\documentclass[12pt]{iopart}
\begin{document}

\title[Hamilton's equations for a fluid membrane: axial symmetry]
{Hamilton's equations for a fluid membrane: axial symmetry}
\author{R  Capovilla\dag ,
J Guven\ddag and E Rojas\S}
\address{\dag\
Departamento de F\'{\i}sica,
 Centro de Investigaci\'on
y de Estudios Avanzados del IPN,
Apdo Postal 14-740, 07000 M\'exico,
D. F.,
MEXICO}
\address{\ddag\
Instituto de Ciencias Nucleares,
 Universidad Nacional Aut\'onoma de M\'exico,
 Apdo. Postal 70-543, 04510 M\'exico, DF, MEXICO}
\address{\S\ Facultad de F\'{\i}sica e Inteligencia
Artificial, Universidad Veracruzana, 91000 Xalapa, Veracruz,
MEXICO}

\begin{abstract}
Consider a homogenous fluid membrane, or vesicle, described by the 
Helfrich-Canham energy, quadratic in the mean curvature.  
When the membrane is axially symmetric, this energy can be viewed as an 
`action' describing the motion of a particle;  
the contours of equilibrium geometries are identified 
with particle trajectories. A novel Hamiltonian formulation of the 
problem is presented which exhibits the following two features: {\it (i)} 
the second derivatives appearing in the action through the mean curvature  
are accommodated in a natural phase space; {\it (ii)} the intrinsic 
freedom associated 
with the choice of evolution parameter along the contour is preserved. As 
a result, the phase space involves momenta conjugate not only to the 
particle position but also to its velocity, and there are constraints 
on the phase space variables. 
This formulation provides the groundwork for a field theoretical 
generalization to arbitrary configurations, with the particle replaced by a 
loop in space.

\end{abstract}

Phospholipid molecules self-assemble in water to form vesicles, or 
membranes \cite{handbook,boal}.
The vesicles are very thin compared to their size so that it is 
sensible to describe them as surfaces of negligible thickness. Moreover, 
the vesicle behaves as a two-dimensional
fluid: there is no resistence to shear so that 
the molecules move freely in the plane of the membrane. 
In a theoretical model of lipid vesicles,
this physical property is realized as invariance under reparametrizations 
of the surface.
The energy thus depends {\it only} on the geometry of this surface.
The leading term in the 
energy is proportional
to the integrated square of the mean curvature, penalizing bending
\cite{canham,helfrich,evans}. The molecular details are essentially irrelevant.

The `shape equation', describing equilibrium 
geometries \cite{helf87,helf89} is 
a fourth order non-linear PDE. 
With axial symmetry, the PDE reduces to a non-linear  ODE which, in turn, 
possesses a first integral \cite{firstintegral,Stress}. 
Indeed, it is possible to interpret the energy as an action describing the motion of 
a particle. Geometrical contours can be identified with particle trajectories.
Building on Deuling and Helfrich's pioneering work in the early seventies \cite{DH}, 
axisymmetric solutions of the shape equation describing an 
isolated vesicle were pretty well understood 
by the mid-nineties (see the reviews \cite{svetina,seifert,lipowsky}). 

Without the symmetry we are less well off.
However, with increasing computer power, impressive results 
can be achieved; 
Monte-Carlo 
 and dynamical triangulation (see {\it e.g.} 
\cite{Gompper,Bowick}) will minimize the energy
for us. In the latter case, the program Surface Evolver was designed with 
exactly this sort of problem in mind \cite{brakke}. 

At this level, the shape equation is consigned to the status of a curiosity.
However, there is a lot of information encoded in the 
shape equation which can be accessed without having to solve it explicitly. 
For example, it is not widely known that the shape equation can be cast as 
a conservation 
law for the stresses prevailing within the membrane \cite{Stress}. These 
stresses are completely geometrical.
They transmit forces. It would be difficult to understand the nature of these 
forces without taking the shape equation apart.
In this respect, a computation scheme to solve  
the shape equation, would be a useful complement to energy minimization.
The mechanical analogue of the 
axially symmetric shape equation is most naturally
formulated as a Hamiltonian initial value problem.
There is no obstacle, in principle, to 
setting up a field theoretical generalization:
instead of a point particle take a  closed loop; 
motion  of the loop will generate a surface.
Unfortunately, 
the existing Hamiltonian approaches to solving the 
axially symmetric shape equation, that use arc-length along the contour as 
a parameter, are tailored very specifically to the symmetry, so they
are not very helpful.

In this paper, we present a novel Hamiltonian formulation 
of the axially symmetric shape equation which takes no  
shortcuts home. It will, however, admit a field 
theoretical generalization with 
the particle replaced by a loop in space \cite{HAM2}. 
This  formulation will involve two key features: 

\vskip0.5em
\noindent {\it (i)}
When axial symmetry is relaxed there is no single  
priveledged  parameter analogous to arc-length along the contour.
The formalism should therefore respect the intrinsic freedom associated with 
the choice of evolution parameter. 

\vskip0.5em
\noindent {\it (ii)}
A point that tends to go unnoticed in the 
axially symmetric is that the action involves not only first derivatives
(velocities) but also second derivatives (accelerations), 
a feature that is somewhat challenging from a Newtonian point of view.
With axial symmetry, the problem  is simply sidestepped 
by introducing the turning angle along the contour ( a velocity)
as an intermediate variable; with respect to this variable, the action 
involves no derivative higher than first. What amounts to the same thing,
only without the sleight of hand, is to introduce the natural phase space that is appropriate for 
the Hamiltonian formulation of a theory based on an action involving accelerations:
introduce momenta not only canonically conjugate to the particle position, but also to
its velocity. 

\vskip0.5em
\noindent  Even if axial symmetry were to be our final goal,
there are benefits to this apparently unnecessarily complicated formalism: 
both the momenta and the constraints possess physical meaning and
Hamilton's equations will evolve physical initial data in a 
remarkably straightforward way.
As we will show the difficulty is in the setup; not in its implementation.

\vspace{1cm}

We model a lipid vesicle as a two-dimensional surface $\Sigma$. The 
surface is
described locally by the embedding ${\bf x} = {\bf X} (u^a )$,
where ${\bf x}$ are local coordinates in space, $u^a = (u^1 , u^2 )$
local coordinates on the surface, and the position functions ${\bf X} (u^a) $ are
three functions of two  variables. We denote by 
${\bf e}_a = \partial_a {\bf X} = \partial {\bf X} / \partial u^a$ the two 
tangent vectors to the surface.
The metric induced on $\Sigma$ is given by their inner product, $g_{ab} = 
{\bf e}_a \cdot {\bf e}_b $. The unit normal ${\bf n}$ to $\Sigma$ is defined
 implicitly by ${\bf e}_a \cdot {\bf n} = 0$, ${\bf n}^2 = 1$.
The extrinsic curvature tensor is $K_{ab} = - {\bf n} \cdot 
\partial_a \partial_b {\bf X}$, and the mean curvature is
$K = g^{ab} K_{ab}$, where $g^{ab}$ is the inverse of the induced 
metric $g_{ab}$.
In terms of the principal curvatures, $\{c_1 , c_2 \}$, we have
$K = c_1 + c_2$.
The intrinsic scalar curvature can be given in terms of the extrinsic
curvature via the 
Gauss-Codazzi equation
as ${\cal R} = K^2 - K^{ab} K_{ab}$; it is twice the Gaussian
curvature $G$, {\it i.e.} ${\cal R} = 2G = 2 c_1 c_2$.

We consider the Helfrich-Canham geometric model, or bilayer
coupling model, for a fluid 
lipid vesicle, with energy 
\begin{equation}
F [{\bf X}] = {\kappa \over 2} \int dA \; K^2 + \beta \int dA \; K
+ \sigma \; A - {\sf P} \; V\,.
\label{eq:model}
\end{equation}
where the constant $\kappa$ is the bending rigidity, $dA = \sqrt{g} d^2 u$ 
denotes the infinitesimal area element on the surface, and $g$ is the
determinant of the induce metric $g_{ab}$.
The constants $\beta$, $\sigma$, ${\sf P}$ are Lagrange multipliers
enforcing the constraints of constant total mean curvature (or
constant area difference between the layers), constant area and
constant enclosed volume $V$, respectively \cite{svetina89}.
A refinement, known as the ADE model, imposes a non-local constraint
involving the square of the area difference \cite{bozic,wiese,miao}. 
Our considerations can be extended to this and other geometrical
models for membranes. 
Note that the volume
can be written as a surface integral:
\begin{equation}
V = {1 \over 3} \int dA \; {\bf n} \cdot {\bf X}\,.
\label{eq:volume}
\end{equation} 
We have not included a term corresponding to the Gaussian bending,
$F_G [{\bf X}] = \kappa_G \int dA {\cal R}$, since 
it is a topological invariant by the Gauss-Bonnet theorem, and, as such, it does not
contribute to the determination of equilibrium configurations. 
The energy (\ref{eq:model})
is invariant under rigid motions, translations and rotations, of the
surface in the ambient space. It also 
possesses a local symmetry: invariance under
reparametrizations. 

The vanishing of the first variation of the  energy (\ref{eq:model}), 
with respect to variations of the position functions ${\bf X} (u^a ) \to 
{\bf X} (u^a ) + \delta {\bf X} (u^a )$, gives the shape equation
\cite{helf87,helf89,Stress,Auxil} 
\begin{equation}
\kappa \left[ -  \nabla^2 K - {K \over 2} ( K^2 - 2 {\cal R} ) \right]
+ \beta {\cal R} + \sigma K - {\sf P} = 0\,,
\label{eq:shape}
\end{equation}
where $\nabla^2$ denotes the surface Laplacian. This 
fourth order non-linear PDE determines the 
equilibrium configurations of lipid vesicles.
There is only one equilibrium condition, whereas naively
one would have expected three. Reparametrization invariance 
informs us that two linear combinations of these three equations, corresponding to
tangential deformations, must vanish identically \cite{Second}. The
only physical deformations are those normal to the surface.

Let us now specialize to
axially symmetric configurations. 
The embedding of an axially symmetric configuration can be written as
\begin{equation}   
{\bf x} = {\bf X} (u^a ) =
{\bf X} ( t , \phi ) =
\left(
\begin{array}{c}  
 R(t) \; \cos \phi , \\
 R(t) \; \sin \phi , \\
 Z(t)
\end{array}
\right)\,,
\end{equation}
where $t$ is an arbitrary  parameter along the contour of the surface at
fixed $\phi$. Any space vector ${\bf V}$ can be written in adapted components
as
\begin{equation}   
{\bf V} (t, \phi ) =
\left(
\begin{array}{c}  
 V_R (t)\; \cos \phi , \\
 V_R (t) \; \sin \phi , \\
 V_Z (t)
\end{array}
\right)\,,
\end{equation}
so that on the plane $\phi=0$ it reduces to a two-dimensional $t$-dependent vector
with components $\{ V_R (t) , V_Z (t)\}$.
The basis adapted to the surface is given by the two tangent 
vectors ${\bf e}_t  =\partial {\bf X} / \partial t = \dot{\bf X}$ and ${\bf 
e}_\phi = 
\partial {\bf X}/ \partial \phi$,
together with the unit normal vector 
\begin{equation}
{\bf n}  ( t , \phi ) = {1 \over N
}
\left(
\begin{array}{c}  
 \dot{Z} (t) \; \cos \phi , \\
 \dot{Z} (t) \; \sin \phi , \\
 - \dot{R}(t)
\end{array}
\right)\,,
\end{equation}
where we introduce the function
\begin{equation}
N =  \sqrt{ \dot{R}^2 + \dot{Z}^2 } \,.
\end{equation}
Note that arclength $l$ along the contour is defined infinitesimally by 
$dl = N \; dt$.
The induced metric and the extrinsic curvature tensor assume the form
\begin{equation}
g_{ab} = \left(
\begin{array}{ll}
N^2
& 0
\\
0 & R^2 
\end{array}
\right)\,, \quad \quad 
K_{ab} = {1 \over N }
\left(
\begin{array}{ll}
\dot{R} \ddot{Z} - \dot{Z} \ddot{R} & 0 \\
0 & R \dot{Z}
\end{array}
\right)\,.
\end{equation}
For the mean  curvature and the scalar curvature it follows that
\begin{equation}
\fl
K = g^{ab} K_{ab} = {  R ( \dot{R} \ddot{Z} - \dot{Z} \ddot{R} ) 
+ N^2 \dot{Z}  \over R N^{3}}\,, \quad \quad \quad
{\cal R} = {2 \dot{Z} ( \dot{R} \ddot{Z} - \dot{Z} \ddot{R} ) \over
R N^4 }\,.
\end{equation}

The Helfrich-Canham  energy (\ref{eq:model}) specialized to a 
axially symmetric configurations, in terms of an arbitrary parameter $t$, 
is 
\begin{equation}
F [{\bf X} ] = 2\pi \int dt \; L 
(R, Z, \dot{R}, \dot{Z}, \ddot{R}, 
\ddot{Z} )\,, 
\label{eq:modelaxi}
\end{equation}
where the Lagrangian function is
\begin{eqnarray}
\fl
L (R, Z, \dot{R}, \dot{Z}, \ddot{R}, 
\ddot{Z} ) 
 &=& {\kappa \over 2} 
{[ R ( \dot{R} \ddot{Z} - \dot{Z} \ddot{R} ) 
+ \dot{Z} N^2 ]^2 
\over
R N^{5}}
+ \beta {R \over N^2 } ( \dot{R} \ddot{Z} - 
\dot{Z} \ddot{R} ) 
+  \beta \dot{Z} \nonumber \\
\fl
&+& \sigma R N 
-
{{\sf P} \over 3} R (R \dot{Z} - Z \dot{R})\,.
\label{eq:lagra}
\end{eqnarray}
We now treat this  energy as an action 
determining the motion of a fictitious particle in the two dimensional
configuration space $\{ R , Z  \}$.
The arbitrary parameter $t$ will play the role of time.
The Lagrangian function involves the squared acceleration of this 
fictitious 
particle; it enters quadratically due to bending, and linearly 
due to the constraint of constant total mean curvature. The only serious  
non-linearity is the dependence
of the Lagrangian  on the velocity $\{ \dot{R} , \dot{Z}\}$.
The factors of $R,N$ ensure that the
action is invariant under reparametrizations of $t$, all that remains 
of the reparametrization invariance of the energy (\ref{eq:model}) once 
we specialize to axially symmetry. The only dependence on $Z$ is through the
volume, if ${\sf P}\ne 0$.

The  axially symmetric
version of the shape equation 
can be obtained either by direct specialization
of the general shape equation (\ref{eq:shape}),  or as the vanishing of the
 Euler-Lagrange derivative of the energy (\ref{eq:modelaxi})  (see the 
Appendix) 
\begin{equation}
{\cal E}_i  = {d^2 \over dt^2 } \left(\partial L \over \partial \ddot{Q}^i 
\right) - {d \over dt } \left(\partial L \over \partial \dot{Q}^i \right)
+ {\partial L \over \partial {Q}^i}\,,
\label{eq:ell}
\end{equation}
where $Q^i = \{ R , Z\}$. Each gives a copy of the 
shape equation.

Two features of the dynamical system defined by the action 
(\ref{eq:modelaxi}) complicate its Hamiltonian formulation:
the energy depends on second derivatives
of the configuration variables, the position functions; 
and because we have chosen to work in an arbitrary parametrization,
there is a local symmetry - reparametrization invariance.
The implications of the latter will become apparent only after we have 
obtained the canonical Hamiltonian. First we must construct the phase space.

The most direct approach to handling the 
presence of the second derivatives $\{ \ddot R, \ddot Z \}$ is to  
extend the phase space: we treat not only the position variables 
$\{ R, Z \}$ but also their velocities $\{ \dot{R} , \dot{Z} \}$ as 
configuration variables, and introduce conjugate momenta for both sets of variables.
(A brief summary of the Hamiltonian formulation of higher derivative systems 
is provided in the Appendix.) The momenta 
$\{ P_R , P_Z \}$
conjugate to the velocities $\{ \dot{R}, \dot{Z}\} $ are, respectively,
\begin{eqnarray}
P_R &=& {\partial L \over \partial \ddot{R}} 
=
- \kappa {\dot{Z} \over N^{5}}
[ R ( \dot{R} \ddot{Z} - \dot{Z} \ddot{R} ) 
+ \dot{Z} N^2 ]
- \beta {R \dot{Z} \over N^2 }\,,
\label{eq:PG1}
\\
P_Z &=& {\partial L \over \partial \ddot{Z}} 
=
 \kappa {\dot{R} \over N^{5}}
[ R ( \dot{R} \ddot{Z} - \dot{Z} \ddot{R} ) 
+ \dot{Z} N^2 ] 
+  \beta {R \dot{R} \over N^2 }\,.  
\label{eq:PG2}
\end{eqnarray}
We note that the vector 
$\{ P_R,P_Z \}$ is directed along the normal
to the contour, and that its bending part is proportional to 
the mean curvature \cite{HAM2}.
Note also that the vector 
$\{ P_R,P_Z \}$  depends at most on second derivatives
of $\{ R, Z \}$.   

The momenta $\{ p_R , p_Z \} $ conjugate to $\{ R , Z\} $ are, respectively,
\begin{eqnarray}
\fl
p_R &=& 
{\partial L \over \partial \dot{R} } - {d \over dt} \left({\partial
L \over \partial \ddot{R}} \right) \nonumber \\
\fl
&=&  -  {5\kappa \dot{R} [ R ( \dot{R} \ddot{Z} - \dot{Z} \ddot{R} ) 
+ \dot{Z} N^2 ]^2  
\over 2 R N^{7} } +  {\kappa  (R \ddot{Z} + 2 \dot{R} \dot{Z} )  [ R ( 
\dot{R} \ddot{Z} - \dot{Z} \ddot{R} ) + \dot{Z} N^2 ]  
\over R N^{5} } 
\nonumber \\
\fl
&-& { 2 \beta R \dot{R} (\dot{R} \ddot{Z} -
\dot{Z} \ddot{R}) 
\over N^4}
+ {\beta R \ddot{Z} \over N^2  }
+ {\sigma R \dot{R} \over N }
+ {{\sf P} \over 3} R Z - \dot{P}_R
\,,
\label{eq:pg1}
\end{eqnarray}
\begin{eqnarray} 
\fl
p_Z &=& {\partial L \over \partial \dot{Z} } - {d \over dt} 
\left({\partial
L \over \partial \ddot{Z} }\right) \nonumber \\
\fl
&=& - {5\kappa \dot{Z} [ R ( \dot{R} \ddot{Z} - \dot{Z} \ddot{R} ) 
+ \dot{Z} N^2 ]^2  
\over 2 R N^{7} }
+ {\kappa (- R \ddot{R} + \dot{R}^2 + 3 \dot{Z}^2 )  [ R ( 
\dot{R} 
\ddot{Z} - \dot{Z} \ddot{R} ) + \dot{Z} N^2 ]  
\over R N^{5} } 
\nonumber \\
\fl
&-& {2 \beta R \dot{Z} (\dot{R} \ddot{Z} -
\dot{Z} \ddot{R}) 
\over N^4 }
- {\beta R \ddot{R} \over N^2 }
+ {\sigma R \dot{Z} \over N } + \beta
- {{\sf P} \over 3} R^2 - \dot{P}_Z
\,.
\label{eq:pg2}
\end{eqnarray}
Despite the unpromising appearance of these expressions,
in  \cite{HAM2} we will see that 
the vector $\{ p_R , p_Z \}$  is 
the projection  of the stress tensor associated with the membrane
along the unit tangent to the contour. There is a direct physical significance
attached. Note that this vector
has a dependence, through the derivatives of
$\{ P_R , P_Z \}$, on the third derivatives of $\{ R , Z \}$.

We have now identified the appropriate phase space for 
the system defined by the energy (\ref{eq:modelaxi}):
the position of a particle in two dimensions $\{ R, Z\}$,
and its conjugate momenta $\{p_R , p_Z \}$, given by (\ref{eq:pg1}),
(\ref{eq:pg2}), together with the velocity $\{\dot{R}, \dot{Z}\}$
and its conjugate momenta $\{ P_R , P_Z \}$, given by (\ref{eq:PG1}), 
(\ref{eq:PG2}). Intuitively, the position is conjugate to its third
derivative; the velocity is conjugate to the second derivative of the
position. As we will see below, however, not all of this phase space is 
accessible: reparametrization invariance will imply constraints.

Our next step is to construct the Hamiltonian on the phase space.
Because the Lagrangian depends on second 
derivatives of $\{ R, Z \}$, 
the definition of the canonical Hamiltonian $H_0$ involves the Legendre 
transformation with respect to the accelerations $\{\ddot{R}, \ddot{Z}\}$ 
as well as the velocities, $\{\dot{R}, 
\dot{Z}\}$, (see the Appendix) as 
\begin{equation}
\fl
H_0 
(P_R, p_R, \dot{R}, R; P_Z, p_Z, \dot{Z}, Z) 
=
P_R \ddot{R} + P_Z \ddot{Z} 
+ p_R \dot{R} + p_Z \dot{Z} - L(R, Z, \dot{R}, \dot{Z}, \ddot{R}, 
\ddot{Z} )\,.
\label{eq:Leg}
\end{equation}
The definition of the momenta $\{ P_R , P_Z\}$ is used to express
the higher derivatives configuration variables  
$\{ \ddot{R}, \ddot{Z}\} $ in terms of 
the phase space 
variables $P_R, P_Z, \dot{R}, \dot{Z}, Z, R$. Unlike a lower order Hamiltonian 
system, the terms $p_R 
\dot{R}$, $p_Z \dot{Z}$ are left alone; they are already in canonical form. 

To facilitate the elimination of $\{ \ddot{R}, \ddot{Z}\}$ in 
(\ref{eq:Leg}), we square
(\ref{eq:PG1}) and (\ref{eq:PG2}), 
defining the momenta $\{ P_R$, $P_Z\}$, and add to give
\begin{equation}
\fl
R^2 ( \dot{R} \ddot{Z} - \dot{Z} \ddot{R} )^2
=
{N^8 \over \kappa^2}  
\left[ \left( P_R  + \kappa {\dot{Z}^2 \over 
N^{3}}
+ \beta {R \dot{Z} \over N^2 } \right)^2 
+  \left( P_Z  - \kappa {\dot{Z} \dot{R} \over 
N^{3}}
- \beta {R \dot{R} \over N^2 } \right)^2 \right]\,.
\end{equation}
It follows that the canonical
Hamiltonian is expressed in terms of the
phase space variables as
\begin{eqnarray}
\fl
H_0 &=&  p_R \dot{R} + p_z \dot{Z} + 
{N^{3} \over 2\kappa R}  
\left[ \left( P_R  + \kappa {\dot{Z}^2 \over 
N^{3}}
+ \beta {R \dot{Z} \over N^2  } \right)^2 
+ \left( P_Z  - \kappa {\dot{Z} \dot{R} \over 
N^{3}}
- \beta {R \dot{R} \over N^2 } \right)^2 \right] \nonumber \\
\fl
&-& {\kappa \dot{Z}^2 \over  2 R N} 
- \beta \dot{Z} 
-  \sigma R N  
+ { {\sf P} \over 3} R (R \dot{Z} - Z \dot{R})\,. 
\label{eq:Hmod}
\end{eqnarray}
This Hamiltonian is quadratic in $\{P_R,  P_Z\}$ and
 linear in $\{ p_R , p_Z \}$. 
 
We have dealt with the first difficulty, the higher order nature of
the system defined by (\ref{eq:modelaxi}), now we face the second one,
the presence of a local symmetry. In this higher
derivative model, the presence of reparametrization invariance implies that
the Hessian of the Lagrangian with respect to the second derivatives
is degenerate, and it is
impossible to invert for the accelerations in terms of their conjugate  
momenta. The Hessian is  
\begin{equation}
H_{ij} = {\partial^2 L \over \partial \ddot{Q}^i \ddot{Q}^j }
= { \kappa R \over 
N^{5} } 
\left(
\begin{array}{ll}
\dot{Z}^2 & - \dot{R} \dot{Z}
\\
- \dot{R} \dot{Z} & \dot{R}^2 
\end{array}
\right)\,,
\end{equation}
with $Q^i = \{ R , Z \}$,
and we see that its determinant vanishes. 
This means that at any value of the parameter $t$
the phase space variables are not all independent, they are
connected by constraints.
The first (or primary) constraint  is easily identified from the 
definition of the higher momenta (\ref{eq:PG1}), (\ref{eq:PG2}) as
\begin{equation}
C = P_R \dot{R} + P_Z \dot{Z} = 0\,.
\label{eq:cp1}
\end{equation}
This is simply the statement  that the vector  
$\{ P_R,P_Z \}$, 
is directed along the normal
to the contour. 
As we will, see this is equivalent to the fact that the tangential component of the acceleration
is gauge: the parametrization we choose will fix this 
component.

The Hamiltonian that generates the motion 
is given by adding this constraint to the canonical Hamiltonian,
\begin{equation}
H = H_0 + \lambda C\,;
\label{eq:hammmo}
\end{equation}
the Lagrange multiplier $\lambda$ is an arbitrary function of $t$
that enforces the constraint (\ref{eq:cp1}).

The Poisson bracket appropriate for this higher derivative model 
is, for any two phase space functions $f,g$ (see the Appendix)
\begin{equation}
\{ f , g \} = 
{\partial f \over \partial \dot{R}}
{\partial g \over \partial P_R } + {\partial f \over \partial R}
{\partial g \over \partial p_R } 
+ {\partial f \over \partial \dot{Z}}
{\partial g \over \partial P_Z } + {\partial f \over \partial Z}
{\partial g \over \partial p_Z } - (f \leftrightarrow g )\,;
\label{eq:PB1}
\end{equation}
the time derivative of a phase space function $f$ is given by the
Poisson bracket with the Hamiltonian (\ref{eq:hammmo})
\begin{equation}
\dot{f} = \{ f , H \} = \{ f , H_0 \} + \lambda \{ f , C \} \,. 
\end{equation}

We have identified a constraint $C$ on the phase space variables.
This is not the whole story, however. Even if 
$C=0$ initially, we are not guaranteed that it continues to hold.
Consistency requires that $C=0$ be preserved by the evolution:
a short calculation gives
\begin{equation}
\dot{C} = \{ C , H_0 \} = - H_0 \,.
\end{equation}
Thus we need to impose the secondary constraint
\begin{equation}
H_0 = 0\,;
\label{eq:s1}
\end{equation}
the canonical Hamiltonian itself must vanish,
the hallmark of reparametrization invariance. Here, it shows up as a 
secondary constraint. Clearly $\dot{H}_0=0$. 
There are no other (tertiary) constraints.
As a constraint, $H_0=0$ specifies the tangential part of the vector 
$\{p_R, p_Z\}$ in terms of the remaining dynamical variables.

The Hamiltonian function (\ref{eq:hammmo}) generating the dynamics 
is a linear combination of two constraints.
Hamilton's equations will reproduce  the equilibrium condition given
by the vanishing of the Euler-Lagrange derivative (\ref{eq:ell}).

The first pair of equations is
\begin{eqnarray}
{ dR \over dt} &=& {\partial H \over \partial p_R } = \dot{R}\,,
\label{eq:h1}
\\
{dZ \over dt} &=& {\partial H \over \partial p_Z } = \dot{Z}\,,
\label{eq:h2}
\end{eqnarray}
since $\{p_R, p_Z \}$ appear in the Hamiltonian only in the combination
$p_R \dot{R} + p_Z \dot{Z}$. 
These equations tell us how the vector $\{ R , Z\}$ evolves; in this 
formalism they are model independent.

The second pair of equations is 
\begin{eqnarray}
{d \dot{R} \over dt} = \ddot{R} &=& {\partial H \over \partial P_R } = 
{N^{3} \over \kappa R}    \left( P_R  + \kappa {\dot{Z}^2 \over 
N^{3}}
+ \beta {R \dot{Z} \over N^2  } \right) + \lambda \dot{R}\,,
\label{eq:h3}
\\
{ d \dot{Z} \over dt} = \ddot{Z} &=& {\partial H \over \partial P_Z } = 
{N^{3}  \over \kappa R}   \left( P_Z  - \kappa {\dot{Z} \dot{R} \over 
N^{3}}
- \beta {R \dot{R} \over N^2 } \right) + \lambda 
\dot{Z} \,. 
\label{eq:h4}
\end{eqnarray}
They tell us how  
$\{\dot R , \dot Z\}$ evolves. 
They involve the Lagrange multiplier $\lambda$ explicitly. 

Just as (\ref{eq:h1}) and (\ref{eq:h2})  
encode the definition of 
the canonical variables $\{ \dot{R}, \dot{Z} \}$
as the time derivatives of $\{ R , Z\}$,
one would expect these equations to encode the definition 
of the momenta $\{ P_R, P_Z\}$ in terms of 
$\{ R , Z\}$, $\{ \dot R , \dot Z\}$ and $\{ \ddot R , \ddot Z\}$.

Let us first express the Lagrange multiplier $\lambda$ in terms of the 
acceleration.
We multiply (\ref{eq:h3}) by $\dot{R}$ and (\ref{eq:h4}) by $\dot{Z}$ and we add.  
Using the constraint (\ref{eq:cp1}), we identify
\begin{equation}
\lambda = { \dot{R} \ddot{R} + \dot{Z} \ddot{Z} \over
N^2 } = {\dot N\over 2 N^2} \,.
\label{eq:lamb}
\end{equation} 
It vanishes in a parametrization by arc-length. Geometrically, it is 
the
affine connection for the planar curve described by $\{ R(t) , Z(t) \}$;
the component of the acceleration tangent to the contour is pure gauge--it
can be chosen arbitrarily. In particular, it can be chosen to vanish.
If the expression (\ref{eq:lamb})
for $\lambda$ is fed back into (\ref{eq:h3}) and (\ref{eq:h4}), we
find that they reproduce the form (\ref{eq:PG1}) and (\ref{eq:PG2}) for
$P_R$ and $P_Z$, respectively. We do, however, have to use the 
primary constraint.

The third pair of equations is
\begin{eqnarray}
\fl
{ d {P}_R \over dt} = - {\partial H \over \partial \dot{R} } &=& 
- p_R - \lambda P_R
- {3  \dot{R} N \over 2 \kappa R} \left(
P_R^2
+ 
P_Z^2 \right) + {P_Z \dot{Z} \over R} 
- {\beta^2 R \dot{R} \over 2 \kappa 
N}
\nonumber \\
\fl
&+& {\beta \over \kappa N}[ P_Z (2\dot{R}^2 +
\dot{Z}^2 ) - P_R \dot{R} \dot{Z} ]
+ \sigma { R \dot{R} \over
N} + {{\sf P} \over 3} R Z\,,
\\
\fl
{ d{P}_Z \over dt} = - {\partial H \over \partial \dot{Z} } &=&  
- p_Z - \lambda P_Z
- {3 \dot{Z} N \over 2 \kappa R}  \left(
P_R^2
+ P_Z^2 \right) + {P_Z \dot{R} \over  R}
-  {2 P_R \dot{Z} \over R} 
- {\beta^2 R \dot{Z} \over 2 \kappa 
N}\,,
\nonumber \\
\fl
&-& {\beta \over \kappa N}[ P_R (\dot{R}^2 +
2 \dot{Z}^2 ) - P_Z \dot{R} \dot{Z} ] + \sigma { R \dot{Z} \over
N } - {{\sf P} \over 3} R^2\,.
\end{eqnarray}
They tell us how the vector $\{P_R,P_Z\}$ evolves.
One would expect these equations 
to encode the definition of the momenta $p_R$ and $p_Z$ given by
(\ref{eq:pg1}) and (\ref{eq:pg2}). 
To show this is not entirely straigtforward.  It is necessary to use the information 
gathered in the previous Hamilton equations, namely the form of $\{ P_R , P_Z \}$ and of $\lambda$.

Finally, the fourth pair of equations is
\begin{eqnarray}
\fl
{d p_R \over dt} = - {\partial H \over \partial R } &=& { N^{3} \over 2 
\kappa R^2 } ( P_R^2 + P_Z^2 ) 
+ { P_R \dot{Z}^2 \over R^2 } - {P_Z \dot{R} \dot{Z} \over R^2 } 
- {\beta^2 N \over 2\kappa } \nonumber \\
\fl
&+& \sigma N  - {{\sf P} \over 3} (2 R \dot{Z} - Z 
\dot{R}) \,,  \\
\fl
{ d p_Z  \over dt} = - {\partial H \over \partial Z } &=& {{\sf P} \over 3}
R\dot{R}\,.
\end{eqnarray}
They tell us how $\{p_R,p_Z\}$ evolve. 
With these equations, we reproduce the vanishing of the Euler-Lagrange
derivative (\ref{eq:ell}).
One sees that the latter of the two equations has the obvious first integral
\begin{equation}
{\cal J} = p_Z - {{\sf P} \over 6} R^2 = \mbox{const.}
\end{equation}
The first integral of the axially symmetric shape equation
\cite{firstintegral,Stress} emerges naturally
within this framework.

The recipe to construct an axially symmetric equilibrium configuration is as follow: 

\noindent Choose initial data:
At $t=0$, take a point on 
the plane, specified by its position vector $\{ R , Z\}$,  
choose a velocity vector $\{ \dot{R} , \dot{Z} \}$
(this encodes the initial direction of the contour);
next choose a vector $\{ P_R , P_Z \}$, 
orthogonal to the velocity (so as to satisfy the primary constraint (\ref{eq:cp1})); 
finally choose the momentum $\{ p_R , p_Z \}$ with a tangential 
component consistent with the secondary constraint $H_0=0$ where
$H_0$ is given by Eq.(\ref{eq:Hmod}). These are our physical degrees of 
freedom. 

\noindent This initial data  set is evolved using Hamilton's equations.
An equilibrium surface contour $\{ R (t) , Z(t)\}$ will be generated. 
The contour itself will not depend on the choice of the lagrange 
multiplier $\lambda$ (or equivalently the choice of the parameter $t$).

\vspace{1cm}

In this paper we have examined
the construction of axially symmetric equilibrium configurations
of a fluid membrane described by the Helfrich-Canham energy
from a Hamiltonian point of view. 
If axial-symmetry were our final aim this would be
a very heavy-handed approach to the problem.
The value of all of this formalism will become apparent 
when we consider the generalization 
to non-axially symmetric configurations \cite{HAM2}. 
This will involve  
stepping up from the Hamiltonian dynamics of a particle to 
the corresponding dynamics of a field
describing a closed curve
in space. The membrane surface will be 
generated by the evolution of this curve.

\ack
We thank Markus Deserno and Martin M\"{u}ller for useful comments.
RC thanks the Aspen Center for Physics for hospitality where part of this
work was carried out. We acknowledge partial support from CONACyT 
grant 44974-F.

\newpage 
\noindent{\bf APPENDIX}

\vspace{.5cm}

For the benefit of the reader unfamiliar with
the Hamiltonian formulation of higher derivative systems, we
consider in this appendix the Hamiltonian description of a toy model:
a particle moving in one-dimension   
described by a Lagrangian of the form $ L = L(q,\dot{q},\ddot{q})$.
The Euler-Lagrange derivative for this Lagrangian is
\begin{equation}
{\cal E} =   
{d^2\over dt^2}\left({\partial L\over \partial \ddot q}\right)
 -
{d\over dt}  \left({\partial L\over \partial \dot q}\right)
+{\partial L\over \partial  q} \,.
\label{eq:EL0}
\end{equation}
The phase space is given by the two conjugate pairs $\{ \dot{q} , P\}$ and
$\{ q , p\}$. 
The momenta conjugate to $\dot{q}$ and $q$ are, respectively,
\begin{eqnarray}
P &=& {\partial L\over \partial \ddot{q}} \,, \\
p &=& {\partial L\over \partial \dot{q}}
- {d\over dt}\,\left({\partial L\over
\partial \ddot{q}}\right)
\,.
\end{eqnarray}
The canonical
Hamiltonian is constructed as the Legendre transformation
with respect to both the acceleration $\ddot{q}$ and the velocity
$\dot{q}$ as
\begin{equation}
H (\dot{q}, P; q, p ) = P \, \ddot q + p \, \dot q - L\,,
\end{equation}
where one uses the definition of the
higher momentum $P$ to express the highest derivative
$\ddot q$ in terms of the phase space variables $P$, $\dot{q}$, and $q$.
The term $p \dot{q}$ in the canonical Hamiltonian is left alone,
since it is already in canonical form.

The Poisson bracket appropriate for this higher derivative model, 
for two arbitrary phase space functions $f,g$, is
\begin{equation}
\{ f , g \} =
{\partial f \over \partial \dot{q}}
{\partial g \over \partial P } + {\partial f \over \partial q}
{\partial g \over \partial p } - (f \leftrightarrow g )\,,
\label{eq:PB}
\end{equation}
and the time derivative of a phase space function is given by this
Poisson bracket with the Hamiltonian
\begin{equation}
{d f\over dt} = \{ f , H \}\,.
\end{equation}
In particular, it follows that the Hamilton equations are
\begin{eqnarray}
{dq\over dt} &=& {\partial H \over \partial p } = \dot{q}\,, \\
{d \dot{q}\over dt} &=& {\partial H \over \partial P }\,, \\
{d {P}\over dt} &=& - {\partial H \over \partial \dot{q} }\,,  \\
{ d {p}\over dt} &=& - {\partial H \over \partial q }\,.
\end{eqnarray}
The first equation identifies the time derivative of $q$
with the canonical variable $\dot q$;
the second equation identifies the form of
the momenta $P$ conjugate to $\dot{q}$; the third equation
identifies the momenta $p$ conjugate to $q$ modulo the definition of $P$. Using the first three
equations, the fourth equation then reproduces the vanishing of the Euler-Lagrange
derivative (\ref{eq:EL0}).

One important special case is given by a Lagrangian {\it linear} in
the acceleration, $L = g (q, \dot{q} ) \ddot{q}$. In this case,
the higher momentum $P = g (q , \dot{q} )$ is independent of $\ddot{q}$ so
that the acceleration cannot be expressed in terms of the canonical
variables. However, it is always possible to add a total derivative
to the Lagrangian, and obtain a Lagrangian that depends at most on
$\dot{q}$ (see {\it e.g.} \cite{FarhiGuthGuven}.


\vspace{1cm}

\section*{References}

\begin{thebibliography}{99}

\bibitem{handbook} Lipowsky R  and Sackmann E  (eds.)  1995 {\it Handbook In 
Biological Physics} vols 1,2 (Amsterdam: Elsevier)
 
\bibitem{boal} Boal D 2002 {\it Mechanics of the Cell} 
(Cambridge: Cambridge U. Press)

\bibitem{canham} Canham P 1970  {\it J. Theor. Biol.} {\bf 26} 61 

\bibitem{helfrich} Helfrich W 1973 {\it Z. Naturforsch.} {\bf C28} 693 

\bibitem{evans} Evans E 1974  {\it Biophys. J. } {\bf 14} 923 


\bibitem{helf87} 
Ou-Yang Z C and  Helfrich W 1987
 {\it Phys. Rev. Lett.} {\bf 59} 2486 

\bibitem{helf89} 
Ou-Yang Z C and  Helfrich W 
1989  {\it Phys. Rev. A} {\bf 39} 5280 


\bibitem{firstintegral} Zheng W and Liu J 1993  {\it Phys. Rev. E}
{\bf 48} 2856 (1993)

\bibitem{Stress} Capovilla R and Guven J 2002 {\it J. Phys. A: Math. and
Gen.}
{\bf 35} 6233 


\bibitem{DH} Deuling H J and Helfrich W 1976 {\it J. Physique (France)}
{\bf 37} 1335 


\bibitem{svetina} Svetina S and 
\u{Z}ek\u{s} B 1996 in
{\it Nonmedical Applications of Liposomes}, eds. D.D. Lasic and
Y. Barenholz (CRC: Boca Raton, FL)

\bibitem{seifert} Seifert U 1997 {\it Adv. in Phys.} {\bf 46} 13 

\bibitem{lipowsky} Lipowsky R 1998 in {\it Encyclopedia of Applied 
Physics} {\bf 23} 199 (Weinheim and New York: VCH Publishers)

\bibitem{Gompper} Gompper G and Kroll D M 1997 {\it J. Phys.: 
Condensed Matter} {\bf 42} 8795 


\bibitem{Bowick} Bowick M and Trassevet A 2001 {\it Phys. Reports}
{\bf 344} 255


\bibitem{brakke} Brakke K A 1992 {\it Experiment. Math.} {\bf 2} 141 

\bibitem{HAM2} 
Capovilla R,   Guven J and Rojas E in preparation


  
\bibitem{svetina89} Svetina S and \u{Z}ek\u{s} B 1989 {\it Eur. 
Biophys. J.} {\bf 17} 101 

\bibitem{bozic} Bozi\u{c} B , Svetina S, \u{Z}ek\u{s} B and  
Waugh R 1992 {\it Biophys. J.} {\bf 61} 963 
 
\bibitem{wiese} Wiese W, Harbich W and Helfrich W 1992 {\it
J. Phys.: Condensed Matter} {\bf 4} 1647 
 
\bibitem{miao} Miao L, Seifert U, Wortis M and  
D\"{o}bereiner H G 1994 {\it Phys. Rev. E} {\bf 43} 5389 
\bibitem{Auxil} 
Guven J 2004 {\it J. Phys. A: Math. and Gen.} {\bf 37} L313 

\bibitem{Second} Capovilla R and Guven J 2004  {\it J. Phys. A: Math.
and Gen.} {\bf 37} 5983 


\bibitem{FarhiGuthGuven} Farhi E, Guth A and Guven J 1990 {\it Nucl. 
Phys. B} {\bf 339} 417 

\end{thebibliography}
\end{document}